\documentclass[11pt]{article}
\usepackage{amssymb}

\newtheorem{theorem}{Theorem}
\newtheorem{lemma}{Lemma}
\newtheorem{definition}{Definition}

\newcommand{\dua}{\mathord{\hbox{\makebox[0pt][l]{\raise .6mm \hbox{$\uparrow$}}$\uparrow$}}}

\def\po{\sqsubseteq}

\def\tr{{\rm{Tr}}}

\def\dom{{\rm{Dom}}}
\def\fix{{\rm{Fix}}}
\def\spec{{\rm{Spec}}}

\def \Scottl    {\leavevmode\raise.1em\hbox{{\rm [\kern-.09em[}}}
\def \Scottr    {\leavevmode\raise.1em\hbox{{\rm ]\kern-.11em]}}}
\def \CTRS      {{\em C\kern-2pt u\kern-1.5pt }{\rm TRS}}
\def \Rm#1{\mbox{\rm #1}}
\def \lsem      {\raise1pt\hbox{\Rm {[\kern-.12em[}}}
\def \rsem      {\raise1pt\hbox{\Rm {]\kern-.12em]}}}
\def \sem#1{\mbox{\lsem$#1$\rsem}}

\title{Quantum Domain Theory - Definitions and Applications}
\author{{\bf Elham Kashefi}\\ \\ {\it Department of Computing} \\ {\it Imperial College London} \\ \\{e.kashefi@ic.ac.uk}}
\date{}

\begin{document}
\pagestyle{plain} \pagenumbering{arabic} \maketitle

\abstract{Classically domain theory is a rigourous mathematical structure to describe denotational semantics for
programming languages and to study the computability of partial functions. Recently, the application of domain
theory has also been extended to the quantum setting. In this note we review these results and we present some new
thoughts in this field.}

\section{Introduction}

In $1930$s the two main models of computation were introduced: $1)$ Turing machines by Alain Turing and $2)$
Lambda calculus by Alonzo Church. The former is the foundation of all existing von-Neumann computers,
computational complexity analysis, and imperative programming languages. The latter is more suitable for the study
of formal methods and functional programming languages. Both directions have been extensively studied in classical
computer science and many equivalent models of computation have been introduced where each new model addresses a
different aspect of information processing.

Quantum computation is traditionally studied via quantum circuit models or in terms of quantum Turing machines,
which fit into the first model of computation \cite{Deutsch85,Deutsch89,Yao93,BV97}. In this approach, one
specifies a concrete recipe for how to build more complicated quantum processes from a few basic building blocks,
and it lays the foundations for computational complexity and the design of new quantum algorithms. In order to
analyse other aspects of quantum computation different models are required. For example, the one-way quantum
computer is a new model for quantum information processing where measurements play a central role \cite{RB02}; it
is a universal model and proven to be polynomially equivalent to the quantum network model. The one-way quantum
computer presents new aspects of quantum processing, such as temporal complexity, that can not be analysed
properly in other models \cite{RBB02}. Recently there have been also some developments in the area of programming
languages, which require models of computation with a higher level of abstraction
\cite{Omer98,Omer00,SZ00,BCS01,Selinger03}.

In this paper we will work within yet another model of computation, namely domain theory. Classically domain
theory is a suitable mathematical structure for descriptions of denotational semantics for programming languages
and to study the computability of partial functions \cite{AJ94,Edalat97}. Recently, the application of domain
theory has also been extended to the quantum setting. Here we review these results and present some new thoughts
in this field. In sections $2$ and $3$ we review the basic concepts of quantum computation and classical domain
theory. In Section $4$ we introduce quantum domain theory and summarise its applications in Quantum Computability,
Quantum Semantics and Quantum Information Theory. Finally Section $5$ contains discussion and indicates further
research directions.

\section{Quantum Computation}

Quantum computation is a rapidly growing cross-disciplinary field which explores the relation between quantum
physics and computation, and is of great importance from both a fundamental as well as technological perspective.
The exciting discovery was that quantum computer is in fact provably more efficient than any classical computer
\cite{Deutsch85,Shor94,Grover96,BBBV97}. One of the key effect leading to this efficiency is the quantum
superposition phenomenon which allows a quantum computer to perform a multitude of different tasks simultaneously
(in parallel). For a general introduction to quantum computing we refer to \cite{NC00}. Here we review some
standard physical background which is required for the discussion of this paper.

The state of a closed quantum physical system which is not interacting with an environment ({\em pure state}) is
described by a unit vector in a Hilbert space, which in Dirac notation is denoted by $|\psi\rangle \in {\cal H}$.
The simplest quantum mechanical system is a qubit, which has a two dimensional Hilbert space, ${\cal
H}_2=\mathbb{C}^2$, as the state space. The state of a composite quantum system (made of $n$ qubits) is described
by the tensor product of state of the component physical systems:
\begin{eqnarray*}
|\psi_1\rangle\otimes\cdots\otimes|\psi_n\rangle \in {\cal H}_{2^n}\, ,
\end{eqnarray*}
where
\begin{eqnarray*}
{\cal H}_{2^n}={\cal H}_2 \otimes \cdots \otimes{\cal H}_2 = \mathbb{C}^{2^n} \, .
\end{eqnarray*}

In Dirac notation, $\langle \psi|$ is used for the dual vector to the vector $|\psi\rangle$ which is a linear
operator over the corresponding Hilbert space defined as following:
\begin{eqnarray*}
\langle \psi |: {\cal H}& \rightarrow& \mathbb{C} \\
|\phi \rangle &\mapsto& \langle \psi| \phi \rangle \, ,
\end{eqnarray*}
where $\langle .|. \rangle$ denotes the inner product of the Hilbert space.

The evolution of a closed quantum system is described by a {\em unitary transformation} \footnote{A transformation
$U$ on a Hilbert space is unitary iff $UU^{\dagger}=U^{\dagger}U=I$.} on the corresponding Hilbert space. The
matrix representations of the quantum operations used in this paper are:
\begin{eqnarray*}
\mbox{Hadamard}\;\;\;\;\; H &=& \frac{1}{\sqrt{2}}\left[ \begin{array}{cc} 1 &\ 1 \\ 1 &\ -1 \end{array}\right], \\
\mbox{Pauli-X}\;\;\;\;\; X &=& \left[ \begin{array}{cc} 0 &\ 1 \\ 1 &\ 0 \end{array}\right], \\
\mbox{Phase}\;\;\;\;\;\;\;\; P &=& \left[ \begin{array}{cc} 1 &\ 0 \\ 0 &\ i \end{array}\right], \\
\mbox{Rotation-$\pi/8$}\;\;\;\;\; T &=& \left[ \begin{array}{cc} 1 &\ 0 \\ 0 &\ e^{i\pi/4} \end{array}\right], \\
\mbox{controlled-Not}\;\;\;\;\; CNOT&=&  \left[ \begin{array}{cccc} 1 &\ 0 &\ 0 &\ 0 \\
0 &\ 1 &\ 0 &\ 0 \\ 0 &\ 0 &\ 0 &\ 1 \\ 0 &\ 0 &\ 1 &\ 0 \end{array}\right]\,.
\end{eqnarray*}
A general unitary transformation $U$ on a finite dimensional Hilbert space can be decomposed into a product of
(one and two)-level unitary matrices where act on $1$ and $2$ qubits. It is known that the set of single qubits
operators {e.g. $H, X, P, T$} and controlled-NOT operator ($CNOT$) are universal. The set of all bounded operators
on a Hilbert space is represented with $B({\cal H})$.

{\em Mixed states} arise when we do not have complete information about the state of the physical system. This is
always the case in experiments, since the system that we are trying to prepare in a pure state interacts with an
uncontrolled environment. A mixed state is a probabilistic mixture of pure states, denoted by
$\{p_i,|\psi_i\rangle\}$ or alternatively by a density matrix
\begin{eqnarray*}
\rho \equiv \sum_i p_i |\psi_i\rangle\langle\psi_i| \, .
\end{eqnarray*}
A density matrix $\rho \in B({\cal H}_{2^n})$ is a hermitian (i.e. $\rho = \rho^\dag$) positive semi-definite
matrix of dimension $2^n\otimes 2^n$ with trace $\mbox{trace}(\rho)=1$. Note that a given pure state
$|\psi\rangle$ can also be represented with the density matrix $|\psi\rangle\langle \psi|$.

The most general operation on quantum states are the transformations of density matrices i.e. linear operators on
operators ({\em super-operator}). The physically allowed super-operators are linear completely positive and
trace-preserving operators, called {\em CP maps} in short. A super-operator $T$, is positive if it sends positive
semi-definite Hermitian matrices to positive semi-definite Hermitian matrices; it is completely positive if
$T\otimes I_d$ is positive, where $I_d$ is the identity operator on $d$-dimensional Hilbert space.

In order to observe a quantum system, a {\em measurement} should be applied. Quantum measurements are described by
a collection ${M_m}$ of {\it measurements operators}. These are operators acting on the state space of the system
being measured. The index $m$ refers to the measurements outcome that may occur in the experiment. If the state of
the quantum system is $\rho$ immediately before the measurement then the probability that the result $m$ occurs is
given by
\begin{eqnarray*}
p(m)=\tr(M_m^{\dagger}M_m \rho) \, ,
\end{eqnarray*}
and the state of the system after the measurement is
\begin{eqnarray*}
\frac{M_m \rho M_m^{\dagger}}{\tr(M_m^{\dagger}M_m \rho)} \, .
\end{eqnarray*}
The measurements operators satisfy the {\it completeness equation},
\begin{eqnarray*}
\sum_m M_m^{\dagger}M_m=I \, .
\end{eqnarray*}

\section{Classical Domain Theory} \label{s-dom}

Domain theory was introduced independently by Scott \cite{Scott70} for the study of denotational semantics and by
Ershow \cite{Ershov72} as a tool for the study of partial computable functions. A complete survey of domain and
its applications can be found in \cite{AJ94,Edalat97}. Domain Theory has been developed towards the following key
applications:

\begin{itemize}
\item A mathematical theory of computation for the semantics of programming languages;
\item A mathematical theory of computation over partial information;
\item An algebraic approach to computability;
\end{itemize}

Informally speaking, a domain is a partially ordered set with added structures to model information processing. In
this pictures of computation, an specific input (output) is represented with a sequence of elements approximating
it. An algorithm is a function from domain of the input to the domain of the output. In order to describe this
model precisely first we introduce the standard basic language of domain theory and all the notations that we will
use later in this paper.

\begin{definition}
A {\em partially ordered set} (poset) is a pair $(P,\po)$, where $\po$ is a binary relation on $P$ such that the
following conditions are satisfied:
\begin{itemize}
\item Reflexibility~. $\forall x\in P:\;\; x\po x$.
\item Transitivity~. $\forall x,y,z \in P:\;\; x\po y \, \& \, y\po z \Rightarrow x \po z $.
\item Anti-symmetry~. $\forall x,y \in P:\;\; x\po y \, \&\, y\po x \Rightarrow x=y$.
\end{itemize}
An element $\bot \in P$ is called a {\em least element} iff $\;\;\forall x \in P:\;\; \bot \po x$.
\end{definition}
It is easy to see that if a poset has a least element, then it is unique.

The poset structure appears in many different fields of computer science and physics and in each context the
ordering , $\po$, is interpreted differently. In this paper, $\po$, refers to a notion of information which will
be described more precisely later. The notion of a sequence of data is captured via the following structures.

\begin{definition}
A subset $A$ in a poset $P$ is called a {\em chain} iff
\begin{eqnarray*}
\forall x,y \in A: \;\; x\po y \;\; \vee \;\; y \po x \, .
\end{eqnarray*}
Assume $A$ is a chain of the poset $P$. An {\em upper bound} of $A$ is an element $u \in P$ such that
\begin{eqnarray*}
\forall x \in A: \;\; x \po l \, ;
\end{eqnarray*}
The {\em least upper bound} of $A$ is denoted by $\sqcup A$.
\end{definition}

Not every chain in a poset has a least upper bound. Adding this property to a poset (chain completeness) will
result in a structure rich enough to model denotational semantics, as we describe later.

\begin{definition} The partially ordered set $P$ is {\em chain-complete} (CCPO) iff all chains $A$ in $P$ have a least upper bound
$\sqcup A$ in $P$.
\end{definition}

We shall be interested in continuous functions:
\begin{definition}
Assume $(P_1,\po_1)$ and $(P_2,\po_2)$ are given posets. A function $f:P_1\rightarrow P_2$ is called {\em
continuous} iff it is :
\begin{itemize}
\item Monotone: $\forall x,y \in P_1: \;\; x\po_1 y \Rightarrow f(x)\po_2 f(y)$.
\item It preserves the least upper bounds of the chains, i.e. for all chains $A$ in $P_1$:
\begin{eqnarray*}
\sqcup_2\{f(x) \,|\, x \in A\} = f(\sqcup_1 A) \, .
\end{eqnarray*}
\end{itemize}
\end{definition}

For a given function $f$, define $f^0$ to be the identity function and $f^{(n+1)}=f\circ f^n$. Now, we can state
the fixed-point theorem which is a canonical tool to construct the mathematical object corresponding to a
recursive definitions.

\begin{theorem} {\bf Knaster-Tarski Fixed-Point Theorem}
Assume $f:P\rightarrow P$ is a continuous function on the chain complete poset $P$ with a least element $\bot$.
Then
\begin{eqnarray*}
\fix f = \sqcup \{ f^n(\bot) \,|\, n\geq 0 \} \, ,
\end{eqnarray*}
defines an element of $P$ which is the least fixed-point of $f$.
\end{theorem}

A similar structure to a chain in a poset is a directed set:
\begin{definition}
A non-empty subset $A \subset P$ of a poset $(P,\po)$ is {\em directed} iff :
\begin{eqnarray*}
\forall x,y \in A \;\; \exists z\in A : x,y \po z \, .
\end{eqnarray*}
\end{definition}
A directed set corresponds to a consistent set of data. We denote by $\sqcup A$ the {\em least upper bound} of a
directed set, if it exists.

\begin{definition}
A partially ordered set in which every directed subset has a least upper bound, is called a {\em domain}.
\end{definition}

The notion of approximation in domain theory is described via the following relation:
\begin{definition}
Assume that $x$ and $y$ belongs to a domain $D$. We say that $x$ is {\em way-below} $y$ or equivalently $x$ {\em
approximates} $y$, denoted by $x\ll y$, iff for every directed subset $A \subset D$:
\begin{eqnarray*}
y\po\sqcup A \;\;\Rightarrow \;\; \exists a\in A : x\po a \, .
\end{eqnarray*}
\end{definition}

A constructive structure for a domain can be introduced via basis elements:
\begin{definition}
A subset $B$ of the domain $D$ is called a {\em basis} iff for each $d\in D$:
\begin{eqnarray*}
A=\{b\in B \,|\, b \ll d\} \;\; \mbox{is directed and} \;\; d=\sqcup A \, .
\end{eqnarray*}
\end{definition}
A domain with a basis is called a {\em continuous} domain and if the basis is also countable the domain is called
an {\em $\omega$-continuous} domain.

The following definitions provide a topological structure for a domain.
\begin{definition}
An {\em open set} $O\subset D$ of the Scott topology of $D$ is a set which satisfies the following conditions:
\begin{itemize}
\item[i)] $x\in O \;\&\; x\po y \Rightarrow y\in O$.
\item[ii)] For any directed subset $A$ of $D$ we have $\sqcup A \in O \Rightarrow \exists x\in A : x\in O$.
\end{itemize}
Dually a {\em closed set} $C\subset D$ is defined with the following conditions:
\begin{itemize}
\item[i)] $x\in C \;\&\; y\po x \Rightarrow y\in C$.
\item[ii)] For any directed subset $A\subset C$ we have $\sqcup A \in C$.
\end{itemize}
\end{definition}
In any continuous domain, subsets $\dua b=\{x\,|\,b\ll x\}$ where $b$ belongs to a given basis of the domain,
forms a basis for the Scott topology.

We denote by $[D\rightarrow D^{\prime}]$ the set of all continuous functions (with respect to the Scott topology)
between two domains $D$ and $D^{\prime}$, which also forms a domain with pointwise ordering:
\begin{eqnarray*}
f\po g \;\;\; \mbox{iff}\;\;\; \forall x \in D: f(x)\po g(x) \, .
\end{eqnarray*}

In summary, in the domain picture of information processing, data are elements of an $\omega$-continuous domain
$D$, and represented as least upper bound of the basis elements. A program is an element of domain of continuous
functions, $[D\rightarrow D]$ and can be represented as the least upper bound of the corresponding basis elements
in $[D\rightarrow D]$. In what follows we review the main applications of domain theory in computability analysis
and denotational semantics. As we show in each scenario a suitable domain will be constructed.

\subsection{Computability Analysis}\label{sub-computable}

There exist two main approaches to computability analysis in the literature. One is the machine-oriented framework
and the other one is the analysis-oriented approach \cite{Weih00}. In the former scenario, the computation is
performed on a certain kind of abstract machine. Whereas in the latter, concepts from classical analysis are
extended to develop a computability theory for real numbers or any other mathematical spaces.

Recently, a new approach to computability has also been developed which is based on domain theory and fits into
the second main framework for computability \cite{WS81,Dig96,Blanck97,ES97}. In his famous article \cite{Scott70},
Scott pointed out the relationship between continuity versus computability. For most purposes to detect whether
some construction is computationally feasible it is sufficient to check that it is continuous, although continuity
is an algebraic condition, which is much easier to handle than computability. We describe briefly how to define
computability via domain theory. In the next section we extend this concept to the quantum setting. We define the
notion of {\em effectively given $\omega$-continuous domain} by putting a proper recursive structure on the
elements of a basis of the domain \cite{Smyth77,Edalat97}.

\begin{definition}
Assume domain $D$ is $\omega$-continuous with a countable basis \\ $B=\{b_0,b_1,b_2,\cdots \}$. We say $D$ is {\em
effectively given with respect to $B$}, if the relation $b_n\ll b_m$ is recursively enumerable \footnote{ A set is
called {\em recursively enumerable} if it is accepted by some Turing machine \cite{Papadimi94}.} in $n$ and $m$.
\end{definition}
The definition of computable elements is:
\begin{definition}
Assume that $D$ is effectively given. An element $x\in D$ is called {\em computable}, if the set $\{n\in
\mathbb{N}\,|\,b_n\ll x\}$ is recursively enumerable.
\end{definition}
We state the following important theorem (without proof) which present a constructive definition of computability.
\begin{theorem}
\cite{ES97} Assume domain $D$ is effectively given, $x \in D$ is computable iff it is the least upper bound of an
effective given chain in the basis $B$ i.e. iff there exists a total recursive function
$f:\mathbb{N}\rightarrow\mathbb{N}$ such that
\begin{eqnarray*}
b_{f(0)}\po b_{f(1)}\po b_{f(2)}\po \cdots \;\;\; \mbox{and}\;\;\; x=\bigsqcup_{n\in \mathbb{N}}b_{f(n)} \, .
\end{eqnarray*}
Moreover, the chain can be chosen to be a $\ll$-chain, i.e. such that
\begin{eqnarray*}
b_{f(0)}\ll b_{f(1)}\ll b_{f(2)}\ll \cdots \, .
\end{eqnarray*}
\end{theorem}

Finally the computability of a function is defined as follows.
\begin{definition}
Assume that domains $D$ and $D^{\prime}$ are effectively given with respect to the basis sets $B$ and
$B^{\prime}$. A continuous function $f:D\rightarrow D^{\prime}$ is called {\em computable}, if the relation
$b^{\prime}_m\ll f(b_n)$ is recursively enumerable in $n$ and $m$.
\end{definition}

\subsection{Denotational Semantics}\label{sub-semantics}

The main problem which gave rise to domain theory was that of describing the meaning of recursive definitions of
objects or data-types \cite{Scott70}. An important result in this direction is the fixed-point theorem.
Traditionally, semantics studies the meaning of programs, mainly in order to be able to state some correctness
properties. The meaning of each phrase in a program is the computation that it describes. There are two main
directions in the area of semantics of programming languages that differ in the areas they are based on:

\begin{itemize}
\item Operational Semantics, basically uses infinite automata, and programs are studied in terms of the steps or
operations by which each program is executed.
\item Denotational Semantics, where programs are interpreted as mathematical \\ functions.
\end{itemize}

Denotational semantics was developed in the early $1970$s by Strachey and Scott \cite{SS71}. They aimed to place
the semantics of programming languages on a purely mathematical basis. Denotational semantics assigns a
mathematical function not only to a complete program but also to every phrase in the language. This approach has
important benefits such as the ability of predicting the behaviour of each program without actually executing it
on a computer or reasoning mathematically about programs, for example to prove that one program is equivalent to
another.

In this subsection we review a denotational semantics introduced by Kozen for probabilistic computation
\cite{Kozen81}. This framework will be the basis of our approach to quantum semantics in the most general case. We
will show that quantum computation over density matrices with completely positive maps, has a similar semantical
structure as probabilistic computation over random variables. First we present some standard basic definitions for
vector spaces \cite{Birkh67,Kozen81}.

\begin{definition}
A subset $\mathbf{P}$ in a vector space $\mathbf{V}$ is called {\em positive cone} iff it satisfies the following
conditions:
\begin{eqnarray*}
\forall x,y \in \mathbf{P} \; \mbox{and positive scalars} \; a,b : ax+by \in \mathbf{P} \\
\forall x \in \mathbf{P} : x,-x \in \mathbf{P} \Rightarrow x=0 \, .
\end{eqnarray*}
\end{definition}
$\mathbf{P}$ induces a partial order on $V$ with the following relation:
\begin{eqnarray*}
x\po_{P} y \;\;\mbox{iff}\;\; y-x \in \mathbf{P} \, .
\end{eqnarray*}

A similar structure to a domain where every directed set has a least upper bound is a {\em lattice} where every
pair of element has a least upper bound. {\em Vector lattices} are the main mathematical structure of the Kozen's
denotational semantics for probabilistic computation.

\begin{definition}
Let $\mathbf{V}$ be a normed vector space and $\mathbf{P}\subset \mathbf{V}$ a positive cone,
$(\mathbf{V},\mathbf{P})$ is called a {\em vector lattice} iff every pair $x,y \in \mathbf{V}$ has a $\po_P$-least
upper bound in $\mathbf{V}$. A vector lattice is called {\em conditionally complete} if every set of elements of
$\mathbf{V}$ with an $\po_P$-upper bound has a least upper bound.
\end{definition}

To partially order a measurable space we will consider Banach lattices.
\begin{definition}
Assume that $\mathbf{B}$ is a normed vector space with norm $\|.\|$, if \\ $(\mathbf{B},\mathbf{P},\|.\|)$ is both
a Banach space and vector lattice such that:
\begin{eqnarray*}
\||x|\|=\|x\| \;\;\; \mbox{and} \;\;\; \forall x,y \in \mathbf{P}: x\po_P y \Rightarrow \|x\| \leq \|y\| \, ,
\end{eqnarray*}
then $\mathbf{B}$ is called a {\em Banach lattice}.
\end{definition}

In the semantics introduced by Kozen for probabilistic computation, programs are interpreted as continuous linear
operators on Banach space of distributions \cite{Kozen81}. In this framework one could work only with the joint
distribution of the program variables instead of dealing directly with variables. Any simple program $P$ maps the
input distributions $\mu$ to the output distribution $P(\mu)$. Kozen has considered a probabilistic WHILE program
over the variables $x_1,\cdots,x_n$. Syntactically, there are five types of statements in the language described
by Kozen \cite{Kozen81}.

\vspace{12pt} \noindent{\em Core syntax of probabilistic WHILE:}
\begin{itemize}
\item simple assignment: $\;\; x_i:=f(x_1,\cdots,x_n)$, where $f:X^n \rightarrow X$ is a measurable function.
\item random assignment: $\;\; x_i:= \mbox{random}$.
\item composition: $\;\; S;T$.
\item conditional: $\;\; \mbox{if}\, B\, \mbox{then}\, S \, \mbox{else}\, T$.
\item while loop: $\;\; \mbox{while}\, B\, \mbox{do}\, S$.
\end{itemize}

Let $(X,M)$ be a measurable space and let $\mathbf{B}=\mathbf{B}(X^n,M^n)$ be the set of measures on the cartesian
product $(X^n,M^n)$. Then $\mathbf{B}$ consists of all possible joint distributions of the program variables
$x_1,x_2,\cdots,x_n$, plus all their linear combinations. Denote by $\mathbf{P}$ the set of all positive measures
and by $ \|.\|$ the total variation norm, then $(\mathbf{B},\mathbf{P},\|.\|)$ is a conditionally complete Banach
lattice \cite{Kozen81}.

Every program $P$ will map a probability distribution into a subprobability measure. This can be extended uniquely
to a linear transformation in $\mathbf{B} \rightarrow \mathbf{B}$. Moreover, this extension will be
$\|.\|$-bounded and therefore continuous. Thus, each program will define a continuous linear operator in
$\mathbf{B} \rightarrow \mathbf{B}$ \cite{Kozen81}.

The space $\mathbf{B}^\prime$ of operators in $\mathbf{B} \rightarrow \mathbf{B}$ forms a Banach space which is
conditionally complete. The partial order on $\mathbf{B}^\prime$ is defined as follows:
\begin{eqnarray*}
S\po T \;\; \mbox{iff}\;\; S(\mu)\po T(\mu) \;\; \mbox{for all} \;\; \mu \in \mathbf{P} \,.
\end{eqnarray*}
Programs will be interpreted as elements of this space. The semantics of the probabilistic WHILE language
,introduced above, is:

\begin{itemize}
\item {\em simple assignment:} If $P$ is the program ``$x_i:=f(x_1,\cdots,x_n)$'' where $f:X^n\rightarrow X$ is a
measurable function, then the meaning of $P$, $\sem{P}$, is the linear operator $P:\mathbf{B}\rightarrow
\mathbf{B}$ such that:
\begin{eqnarray*}
P(\mu)=\mu\circ F^{-1}\, ,
\end{eqnarray*}
where $F:X^n\rightarrow X^n$ is the measurable function
\begin{eqnarray*}
F(a_1,\cdots,a_n)=(a_1,\cdots,a_{i-1},f(a_1,\cdots,a_n),a_{i+1},\cdots a_n) \, .
\end{eqnarray*}
Since $f$ is measurable, so is $F$, thus $\mu\circ F^{-1}$ is indeed a measure.
\item {\em random assignment:} If $P$ is the program ``$x_i:= \mbox{random}$'' then the meaning of $P$, $\sem{P}$, is the linear operator $P:\mathbf{B}\rightarrow \mathbf{B}$ such that:
\begin{eqnarray*}
P(\mu)(B_1\times\cdots\times B_n)=\mu(B_1\times\cdots,B_i,X,B_{i+1},\cdots B_n)\rho(B_i)\, ,
\end{eqnarray*}
where $\rho$ is a fixed distribution.
\item {\em composition:} The meaning of the program ``$S;T$'' is the composition of operators $\sem{S}\circ \sem{T}$.
\item {\em conditional:} Let $\mu_B$ denote the measure $\mu_B(A)=\mu(A\cap B)$. The conditional test checks the membership
of $x_1,\cdots,x_n$ in $B$, which will occur with probability $\mu(B)$ and hence $S$ will be executed on the
conditional probability distribution $\mu_B/\mu(B)$. Similarly, with probability $\mu(\neg B)$ the program $T$
will be executed on $\mu_{\neg B}/\mu(\neg B)$. Formally, the semantics of the program ``$\mbox{if}\, B\,
\mbox{then}\, S \, \mbox{else}\, T$'' is the linear operator \\ $P:\mathbf{B}\rightarrow \mathbf{B}$ such that:
\begin{eqnarray*}
A &\mapsto& \mu(B)S(\mu_B/\mu(B))(A)+\mu(\neg B)T(\mu_{\neg B}/\mu(\neg B))(A) \\
&=& (S(\mu_B)+T(\mu_{\neg B}))(A) \, ,
\end{eqnarray*}
which can be written as $S\circ e_B + T\circ e_{\sim B}$ where $e_B$ is the operator $e_B(\mu)=\mu_B$ and $+$ is
addition in $B^{\prime}$.
\item {\em while loop:} The meaning of the program ``$\mbox{while}\, B\, \mbox{do}\, S$'' is equivalent to the program
\begin{eqnarray*}
\mbox{if}\,\neg B\,\mbox{then}\,I\,\mbox{else}\, S; \mbox{while}\, B\, \mbox{do}\, S \, ,
\end{eqnarray*}
therefore the meaning of a ``while statement'' must be a solution of
\begin{eqnarray*}
W=e_{\sim B}+W\circ P \circ e_B \, .
\end{eqnarray*}
Using well established techniques one can solve the above equation to drive the following solution. The meaning of
a ``while statement'' is the fixed-point of the affine transformation $\tau: \mathbf{B}^\prime \rightarrow
\mathbf{B}^\prime$ defined by
\begin{eqnarray*}
\tau(W)=e_{\sim B}+W\circ S \circ e_B \, ,
\end{eqnarray*}
which is equal to
\begin{eqnarray*}
\tau^n(0)=\sum _{0 \leq k \leq n-1} e_{\sim B}\circ (S \circ e_B)^k \, .
\end{eqnarray*}
\end{itemize}

\section{Quantum Setting}

In this section we present some of the applications of domain theory in the framework of quantum computation. In
the first subsection we study the domain computability for quantum computation. Subsequently a denotational
semantics for quantum computation is presented. Finally we review a recent work on information aspects of quantum
domain theory by Coecke and Martin \cite{CM02}. By introducing a domain framework for quantum computation we aim
to address different aspects of information processing which has not yet been studied in other existing models of
quantum computation.

\subsection{Computability Analysis}

The Church-Turing thesis is about classical computability, (i.e. the computability which is defined based on a
computing machine which obeys classical mechanics). Hence, it might be thought that quantum mechanical computing
can violent the Church-Turing thesis. However, Deutsch \cite{Deutsch85} and Jozsa \cite{Jozsa91} discussed this
problem and showed that the class of functions computable by a deterministic quantum Turing machine, is equal to
the class of recursive functions (computable by a classical Turing machine). Ozawa extended this argument to the
probabilistic quantum Turing machine \cite{Ozawa98}. He also distinguished the notation of measurability from
computability to answer the problem that has been alleged by Nielsen in \cite{Nielsen97}.

Apart from these few discussions, there have been no further attempts in this direction. We believe, by
introducing a rigourous framework for quantum computability, we can address more interesting questions.
Furthermore, quantum domain theory provides a topological structure for quantum computation that can be useful for
the study in other fields of quantum computation.

To develop a computational model to analyse quantum computability, it would be enough to consider a model for a
Hilbert space. Different effective structures for metric spaces can be found in the literature. We use the domain
of the closed balls \cite{WS81,EH97} to introduce a model for quantum pure states and the power domain of the
former domain \cite{JP89, Edalat97, Martin01} will capture the quantum mixed states. Most of the definitions of
this subsection have been already appeared in \cite{EH97, Edalat97} under the theory of computability for Metric
spaces. We rephrase these results in order to suit our purposes of defining a mathematical foundation for quantum
computability.

\vspace{12pt} \noindent{\bf Pure quantum states}\vspace{12pt}

An standard way to construct a partially ordered set for a given metric space $(X,d)$ is based on ordering of the
set of closed balls \cite{Heckmann99}. Define a closed ball $C(x,r)$ of given metric space $(X,d)$ with $x \in X$
and $r \in \mathbb{R}$ to be the following set:
\begin{eqnarray*}
C(x,r)=\{y\in X\,|\,d(x,y)\leq r\} \, .
\end{eqnarray*}

The Hilbert space ${\cal H}$ of the quantum pure state is a metric space by virtue of the metric induced by the
standard scalar product. Denote the poset of all closed balls of $\mathcal{H}$ by $C\mathcal{H}$ with the
following partial order:
\begin{eqnarray*}
C(|\phi\rangle,r)\po C(|\psi\rangle,s)\;\;\;\mbox{iff}\;\;\;C(|\phi\rangle,r)\supseteq C(|\psi\rangle,s).
\end{eqnarray*}
This relation reflects a natural notion of information: $C(|\phi\rangle,r)\po C(|\psi\rangle,s)$ can be read as
the statement that $(|\phi\rangle,r)$ has less information than $(|\psi\rangle,s)$. The quantum pure state
$|\phi\rangle \in \mathcal{H}$ can be identified with the maximal closed ball $C(|\phi\rangle,0) \in
C\mathcal{H}$, i.e.~the maximal element of the poset $C\mathcal{H}$ is in one-to-one correspondence with
$\mathcal{H}$. The following results from \cite{EH97} prove that the poset $C\mathcal{H}$ has the required
structure for the foundation of a computational model.

\begin{theorem} \cite{EH97} Let $B$ be a dense subset of a separable Hilbert space $\mathcal{H}$. Then
$B\times\mathbb{Q^{+}}$ is a basis of $C\mathcal{H}$ where $\mathbb{Q}^{+}$ is the set of all non-negative
rational numbers.
\end{theorem}

There are many different choices for a dense subset of $\mathcal{H}$. Any universal set of quantum gates defines a
different dense subset of quantum states of a Hilbert space $\mathcal{H}$. To see this fact consider a discrete
set of universal quantum gates, $\mathcal{S}$, (e.g. $H, X, P, T, CNOT$), therefore any unitary operator on
$\mathcal{H}$ can be approximated by combination of elements in $\mathcal{S}$. In other words a universal set of
gates is a dense subset of the set of all unitary operators on $\mathcal{H}$. Denote by $<\mathcal{S}>$ the set of
all finite combination of elements of $\mathcal{S}$. The following lemma gives a dense subset of $\mathcal{H}$.

\begin{lemma}
The image of $<\mathcal{S}>$ on state $|0\rangle \in \mathcal{H}$ is a dense subset of $\mathcal{H}$.
\end{lemma}

\begin{theorem} \cite{EH97} The poset of the closed balls of a separable Hilbert space, ordered by reversed inclusion,
is an $\omega$-continuous domain.
\end{theorem}

It is easy to see that the way-below relation is nothing but
\begin{eqnarray*}
C(|\phi\rangle,r)\ll(|\psi\rangle,s)\;\;\;\mbox{iff}\;\;\;C(|\phi\rangle,r)\supset C(|\psi\rangle,s).
\end{eqnarray*}

The embedding of $\mathcal{H}$ into $C\mathcal{H}$ is defined with the following function:
\begin{eqnarray*}
e_P:\mathcal{H}&\rightarrow& C\mathcal{H} \\
|\phi\rangle&\mapsto&(|\phi\rangle,0) \, .
\end{eqnarray*}
Clearly, the elements of $C\mathcal{H}^{+}=\{(|\phi\rangle,0)\,|\,|\phi\rangle \in \mathcal{H}\}$ are the maximal
elements of $C\mathcal{H}$. Following the definitions of Subsection \ref{s-dom}, we can introduce a topological
structure for $C\mathcal{H}$. It is easy to check that for any given element $(|\phi\rangle,r)\in C\mathcal{H}$ we
have:
\begin{eqnarray*}
e_{P}^{-1}(\dua(|\phi\rangle,r))=O(|\phi\rangle,r)\, ,
\end{eqnarray*}
where $O(|\phi\rangle,r)$ is the open ball with centner $|\psi\rangle$ and radius $r$. The subsets
$\dua(|\phi\rangle,r)$ form a basis of the Scott topology on $C\mathcal{H}$, while the open balls
$O(|\phi\rangle,r)$ are a basis for metric topology on $\mathcal{H}$. Hence, $e_P$ is a topological embedding,
which makes $\mathcal{H}$ homomorphic to the subspace of maximal elements of $C\mathcal{H}$.

The $\omega$-continuity of $C\mathcal{H}$ introduces an effective structure along the lines of Subsection
\ref{sub-computable}. The homomorphism between $\mathcal{H}$ and maximal elements of $C\mathcal{H}$, derives an
effective structure for $\mathcal{H}$ and hence it provides a computational framework for quantum pure states. In
a similar way to the Subsection \ref{sub-computable} we can define a computable pure state as follows.

\begin{definition}
An quantum pure state $|\psi\rangle$ is called {\em computable}, if its domain image
$e_P(|\psi\rangle)=(|\psi\rangle,0)$ is computable in $C\mathcal{H}$, i.e. iff the set $\{n\in
\mathbb{N}\,|\,b_n\ll (|\psi\rangle,0) \}$ is recursively enumerable (where $\{b_n\}$ are elements of the basis
$\mathcal{B}_{C\mathcal{H}}$).
\end{definition}

\vspace{12pt} \noindent{\bf Mixed quantum states}\vspace{12pt}

The Gleason theorem provides a correspondence between density matrices and probability measures on $\mathcal{H}$
\cite{Gleason57}.

\begin{theorem}\cite{Gleason57}\label{t-Gleas} Let $\mu$ be a probability measure on the closed subspaces of a separable Hilbert space
$\mathcal{H}$ of dimension at least three. There exists a positive semi-definite self-adjoint operator $T$ of the
trace class (density matrix) such that for all closed subspaces $A$ of $\mathcal{H}$
\begin{eqnarray*}
\mu(A)=\tr(T P_A) \, ,
\end{eqnarray*}
where $P_A$ is the orthogonal projection of $\mathcal{H}$ onto $A$.
\end{theorem}
Therefore, to present a computational framework for mixed states, it is enough to construct such a framework for
probability measure on $\mathcal{H}$. To this end we need the following notations and results
\cite{Heckmann94,Edalat97,Alvarez00,Martin01}.

The domain of probability measures will be defined in terms of continuous valuation functions, a finite measure
which is defined on open subsets of a topological space \cite{Birkh67,Heckmann96,Edalat97}.

\begin{definition}
Assume that $X$ is a topological space. A function $\nu$ from open sets of $X$ to non-negative real number,
$\mathbb{R}^{+}$, is called a {\em continuous valuation} function iff the following conditions are satisfied:
\begin{itemize}
\item Strictness. $\nu(\emptyset)=0$;
\item Monotonicity. $A\subseteq B \Rightarrow \nu(A)\leq \nu(B)$;
\item Modularity. $\nu(A\cup B)+\nu(A\cap B)=\nu(A)+\nu(B)$;
\item Continuity. whenever $\mathcal{I}$ is a directed subset of open sets
(with respect to $\subseteq$), \\ $\nu(\bigcup \mathcal{I})=\sup_{A\in\mathcal{I}}\nu(A)$.
\end{itemize}
A continuous valuation on an $\omega$-continuous domain is a continuous valuation on its Scott topology.
\end{definition}

\begin{definition}\cite{JP89}
Assume that $X$ is a topological space. The {\em probabilistic power domain} $PX$ of $X$ consists of all
continuous valuations $\nu$ on $X$ with $\nu(X)\leq 1$, ordered pointwise, i.e.
\begin{eqnarray*}
\mu \po \nu \;\;\;\mbox{iff}\;\;\;\mu(O)\leq\nu(O) \; \mbox{for all open sets in} \; X \, .
\end{eqnarray*}
\end{definition}

Simple valuation functions provide a basis for probabilistic power domain.

\begin{definition}\cite{JP89}
For any point $x \in X$ the {\em point valuation}, $\delta_x$, is defined as follows:
\begin{eqnarray*}
\delta_x(O) = \left\{
\begin{array}{ll}
1 &\ \mbox{if}\ x\in O\\
0 &\ \mbox{if}\ x \not\in O
\end{array}
\right.
\end{eqnarray*}
A finite linear combination of point valuations i.e. $\sum_{i=1}^{n} r_i\delta_{x_i}$ with $x_i \in X$ and
positive rational numbers $r_i$ satisfying $\sum_{i=1}^{n} r_i \leq 1$, is called a {\em simple valuation}.
\end{definition}

\begin{theorem}\cite{JP89}
The probabilistic power domain of an $\omega$-continuous domain is also $\omega$-continuous with a basis of simple
valuation.
\end{theorem}

Now, we can introduce the domain of quantum mixed states. The set of all closed subspaces of $\mathcal{H}$ is the
$\sigma$-algebra, $\mathcal{M}$, of the measurable sets. Let $\mathbf{M}(\mathcal{H})$ denote the set of all
probability measures on $\mathcal{H}$.  Based on Gleason theorem (Theorem \ref{t-Gleas}), a mixed state can be
considered to be an element of $\mathbf{M}(\mathcal{H})$. We embed $\mathbf{M}(\mathcal{H})$ into the
probabilistic power domain $PC\mathcal{H}$ of the closed ball domain $C\mathcal{H}$, which forms an
$\omega$-continuous domain.

The set of maximal elements of $PC\mathcal{H}$ contains all valuations $\nu$ such that:
\begin{eqnarray*}
\nu(O)=1 \;\;\; \mbox{for all open subsets}\;\;\; O\in C\mathcal{H}^{+} \, .
\end{eqnarray*}
The embedding of $\mathbf{M}(\mathcal{H})$ into $PC\mathcal{H}$ is defined with the following function:
\begin{eqnarray*}
e_M:\mathbf{M}(\mathcal{H})&\rightarrow& PC\mathcal{H} \\
\mu&\mapsto&\mu\circ e_P^{-1} \, .
\end{eqnarray*}

The following result from \cite{Edalat97} provides the correspondence between $\mathbf{M}(\mathcal{H})$ and
$PC\mathcal{H}^{+}$:
\begin{theorem} \cite{Edalat97}
The space $M\mathcal{H}$ is homomorphic with the space of maximal elements of the $\omega$-continuous domain
$PC\mathcal{H}$. These maximal elements are characterised by $\nu(C\mathcal{H}^{+})=1$. Every mixed state on
$\mathcal{H}$ can be obtained via this homomorphism as the least upper bound of an increasing chain of simple
valuations on $C\mathcal{H}$.
\end{theorem}

Similar to the case of pure states, we define the computability of a mixed state via the computational framework
of $PC\mathcal{H}$.

\begin{definition}
A quantum mixed state $\rho$ is called {\em computable}, if its corresponding measure $\mu_{\rho} \in
\mathbf{M}(\mathcal{H})$ is computable i.e. if the domain element $e_M(\mu)$ is computable in $PC\mathcal{H}$,
i.e. iff the set $\{n\in \mathbb{N}\,|\,b_n\ll e_M(\mu) \}$ is recursively enumerable (where $\{b_n\}$ are
elements of the basis $\mathcal{B}_{PC\mathcal{H}}$).
\end{definition}

The process of quantum computation over a pure state is described with a unitary operator and over a mixed state
is described with a CP map. As we explained before, in a domain-picture of computation, programs are functions
from the domain of the input to the domain of the output. The set of all continuous function forms the domain of
operators. This is exactly the same in the case of quantum computation.

\vspace{12pt} \noindent{\bf Unitary Operators} \vspace{12pt}

We denote by $[C\mathcal{H}\rightarrow C\mathcal{H}]$, the domain of the all continuous functions on
$C\mathcal{H}$ with pointwise ordering. Every unitary operator $U:\mathcal{H}\rightarrow\mathcal{H}$ has a
Scott-continuous extension to the domain of $[C\mathcal{H}\rightarrow C\mathcal{H}]$, i.e.~ there exists a
Scott-continuous function $\tilde{U}$ in $[C\mathcal{H}\rightarrow C\mathcal{H}]$ such that
\begin{eqnarray*}
\tilde{U}(C(|\phi\rangle,0)) = C(U|\phi\rangle,0) \;\;\; \mbox{for all} \;\;\; |\phi\rangle \in \mathcal{H} \, ,
\end{eqnarray*}
and it is explicitly given by
\begin{eqnarray*}
\tilde{U}(C(|\phi\rangle,r)) = C(U|\phi\rangle,r) \, .
\end{eqnarray*}
The following lemma shows that the $\tilde{U}$ is well-defined.
\begin{lemma}
Let $U$ to be a unitary operator on $\mathcal{H}$. The extension function $\tilde{U}$ (defined above), maps closed
balls in $C\mathcal{H}$ to another closed ball.
\end{lemma}

We define the computability of a unitary function via domain theory with:
\begin{definition}
A unitary function $U:\mathcal{H}\rightarrow\mathcal{H}$ is {\em computable} iff its extension, $\tilde{U}$, is
computable in $[C\mathcal{H}\rightarrow C\mathcal{H}]$ (in the terms of Subsection \ref{sub-computable}).
\end{definition}

\vspace{12pt} \noindent{\bf CP Maps}\vspace{12pt}

For simplicity, we denote by $T_\mu$ the corresponding operator for a given measure $\mu$ which is derived from
Theorem \ref{t-Gleas}:
\begin{eqnarray*}
\forall \mu \;\; \exists T : \mu(A)=\tr(T P_A) \;\; \mbox{for all closed subspace} \;\; A \, .
\end{eqnarray*}

A CP map $\mathrm{A}$ is an operator over $B(\mathcal{H})$ and can also be considered as a function in
$\mathbf{M}(\mathcal{H})\rightarrow \mathbf{M}(\mathcal{H})$ (from Gleason theorem). Denote by
$[PC\mathcal{H}\rightarrow PC\mathcal{H}]$ the domain of all continuous functions on $PC\mathcal{H}$ (with
pointwise ordering).

Every CP map $\mathrm{A}:\mathbf{M}(\mathcal{H})\rightarrow\mathbf{M}(\mathcal{H})$ has a Scott-continuous
extension to the domain of $[PC\mathcal{H}\rightarrow PC\mathcal{H}]$, i.e. there exists a Scott-continuous
function $\tilde{\mathrm{A}} \in [PC\mathcal{H}\rightarrow PC\mathcal{H}]$ such that for every probability measure
$\mu \in \mathbf{M}(\mathcal{H})$ we have:
\begin{eqnarray*}
\tilde{\mathrm{A}}(\mu) = \mathrm{A}(\mu) \, .
\end{eqnarray*}
The extension function $\tilde{\mathrm{A}}$ for a continuous valuation function $\nu \in PC\mathcal{H}$ is
explicitly given by:
\begin{eqnarray*}
\tilde{\mathrm{A}}(\nu)(B) = \tr(T_{\mathrm{A}(\nu)} P_B) \;\;\; \mbox{where $B$ is a closed subspace of} \;\;\;
\mathcal{H} \, .
\end{eqnarray*}
The following lemma shows that the above definition is well-defined:
\begin{lemma}
Any CP map $\mathrm{A}$ on $\mathbf{M}(\mathcal{H})$, maps continuous valuation functions to another continuous
valuation function.
\end{lemma}

We define the computability of a CP map function via domain theory with:
\begin{definition}
A CP map $\mathrm{A}:\mathbf{M}(\mathcal{H})\rightarrow\mathbf{M}(\mathcal{H})$ is computable iff its extension,
$\tilde{\mathrm{A}}$, is computable in $[PC\mathcal{H}\rightarrow PC\mathcal{H}]$.
\end{definition}

\vspace{12pt} \noindent{\bf Quantum Measurements}\vspace{12pt}

At the end of computation a measurement operator will be applied. A measurement can be viewed as a CP map which
takes a density matrix (the final state) to another density matrix (the probabilistic mixture of the outcomes).

Assume ${M_m}$ is a collection of measurement operators. The corresponding measurement of this collection can be
considered as a CP map over $B(\mathcal{H})$ :
\begin{eqnarray*}
\mathrm{M}:B(\mathcal{H})&\rightarrow& B(\mathcal{H})\\
\rho&\mapsto&\frac{M_m\rho M_m^\dag}{\tr(M_m^\dag M_m \rho)} \, .
\end{eqnarray*}
Hence, the extension function and computability can be also defined exactly in the same way that we defined before
for a given CP map.

\subsection{General Setting}

In this subsection we present a denotational semantics for quantum computation using domain theory, which could be
considered as a foundation for designing a functional programming language for quantum computation. The resent
literature contains several proposals for quantum programming languages. The first contribution in this direction
is the Knill's paper on QRAM model \cite{Knill96}. The other attempts to define a true quantum programming
language are two imperative languages. The first approach by \"{O}mer \cite{Omer98,Omer00} has a C-like syntax,
while a second proposal by Sanders and Zuliani \cite{SZ00} is based on Dijkstra's guarded-command language. A
similar approach to the work of this subsection has been developed independently by Selinger \cite{Selinger03}. He
has presented the first functional programming language and discussed the denotational semantics of his proposed
language. Our work is based on the Kozen's semantics for probabilistic computation \cite{Kozen81}.

We aim to develop a denotational semantics for a basic programming language, called Quantum WHILE. In this
approach, we show how to define the mathematical object corresponding to the language constructors. We will
consider a simple quantum computational machine with quantum memory registers. To develop the proper foundation
for quantum semantics, in the most general setting, we consider density matrices and  CP maps. Aharanov, Kitaev
and Nisan in \cite{AKN98} introduced the first computational model based on mixed state where possible operators
are represented by CP maps. We show in this subsection that the same structure of the classical probabilistic
semantics which has been introduced by Kozen \cite{Kozen81} can also capture the semantics of quantum computation.

To follow the procedure introduced by Kozen \cite{Kozen81}, we define a measurable space
\\ $(\mathcal{X}^n,\mathcal{M}^n)$ with the set of all measures
$\mathbf{B}=\mathbf{B}(\mathcal{H}_1\otimes\cdots\otimes \mathcal{H}_n, \mathcal{M}^n)$ such that the set of all
probability measures in this space is in correspondence with the set of all density matrices over ${\cal
H}_1\otimes\cdots\otimes{\cal H}_n$. In this way, input to a quantum program $P$ is represented by a probability
measure $\rho \in \mathbf{B}$ which is the same as the corresponding density matrix of all input pure states
$|\phi_1\rangle\otimes\cdots\otimes|\phi_n\rangle$ in $\mathcal{H}_1\otimes\cdots\otimes \mathcal{H}_n$.

Let $\mathcal{H}=\mathcal{H}_1\otimes\cdots\otimes \mathcal{H}_n$ denote the Hilbert space spanned by all the
quantum variables which are involved in the computation. Define $\mathcal{X}_i$ to be the set of all unit vectors
in $\mathcal{H}_i$ and $\mathcal{X}$ to be the set of all unit vectors in $\mathcal{H}$. The set of all closed
subspaces of $\mathcal{H}$ is the $\sigma$-algebra, $\mathcal{M}$, of the measurable sets. Gleason theorem
determines all measures on $\mathcal{M}$ and shows a correspondences between operators in
$\mathrm{B}(\mathcal{H})$ and measures on $\mathcal{M}^n$, we use interchangeably any of the two notions of
measure and operator. In the same way as the classical case, the set of positive measures (positive self-adjoint
operators) $\mathbf{P} \subset \mathbf{B}$ is the positive cone of the measure space $\mathbf{B}$. The definition
of ordering of measures is defined as follows
\begin{eqnarray*}
\mu \sqsubseteq \nu \;\;\mbox{iff}\;\; \nu -\mu \in \mathbf{P} \, .
\end{eqnarray*}

In semantics of the general setting for quantum computation, each program will define a CP map. For simplicity, in
what follows the symbol $P$ refers to both a program and the corresponding CP map. A quantum program $P$ maps
distributions $\mu$ on $(\mathcal{X},\mathcal{M})$ to distribution $P(\mu)$ on $(\mathcal{X},\mathcal{M})$, or
equivalently, maps a density matrix $\mu$ on $\mathcal{H}$ to the density matrix $P(\mu)$.

For the completeness of the discussion we will give the full semantics of the quantum WHILE language in the
general setting. The syntax of this Language is the same as the syntax of the classical probabilistic WHILE
language (Subsection \ref{sub-semantics}). The only difference is that instead of ``random assignment'' we have
``quantum measurement''.

\begin{itemize}
\item {\em simple assignment}: If $P$ is the program ``$x_i:=f(x_1,\cdots,x_n)$'' where $f:\mathcal{X}\rightarrow
\mathcal{X}_i$ is a measurable function, then the meaning of $P$ is the following CP map:
\begin{eqnarray*}
\mu &\mapsto& P(\mu) \\
P(\mu)&=&\mu\circ F^{-1}\, ,
\end{eqnarray*}
where $F:\mathcal{X}\rightarrow \mathcal{X}$ is the measurable function
\begin{eqnarray*}
F(a_1,\cdots,a_n)=(a_1,\cdots,a_i,f(a_1,\cdots,a_n),a_{i+1},\cdots a_n) \, .
\end{eqnarray*}

\item {\em measurement assignment}: If $P$ is the program ``$x_i:= \mbox{measure}$'' then the meaning of $P$ is the
following CP map:
\begin{eqnarray*}
\mu &\mapsto& P(\mu) \\
P(\mu)(B_1\otimes\cdots\otimes B_n)&=& \mu(B_1\otimes\cdots\otimes B_{i-1}\otimes\mathcal{X}_i\otimes
B_{i+1}\otimes\cdots\otimes B_n) \\
& & T_{\rho}(B_i)\, ,
\end{eqnarray*}
where $T_{\rho}$ is a fixed distribution corresponding to the measurement process. To be more precise, assume that
the collection $\{M_m\}$ describes the quantum measurement that has been applied, then
\begin{eqnarray*}
\rho= \frac{M_m \mu M^{\dag}_m}{\mbox{tr}(M^{\dag}_m M_m \mu)} \, ,
\end{eqnarray*}
and $T_{\rho}$ is the corresponding measure obtained from Gleason theorem.

\item {\em composition}: The meaning of the program ``$S;T$'' is the functional composition of the CP maps $T$ and $S$, $T\circ S$.
\item {\em conditional}: The semantics of the program ``$\mbox{if}\, B\, \mbox{then}\, S \, \mbox{else}\, T$'' is the
CP map
\begin{eqnarray*}
S\circ e_B + T\circ e_{\sim B} \, ,
\end{eqnarray*}
where $e_B$ is the CP map $e_B(\mu)=\mu_B$.
\item {\em while loop}: The meaning of the program ``$\mbox{while}\, B\, \mbox{do}\, S$'' is the fixed-point of
the affine transformation $\tau : \mathbf{B}^\prime \rightarrow \mathbf{B}^\prime$ defined by
\begin{eqnarray*}
\tau(W)=e_{\sim B}+W\circ S \circ e_B \, ,
\end{eqnarray*}
which is equal to
\begin{eqnarray*}
\tau^n(0)=\sum _{0 \leq k \leq n-1} e_{\sim B}\circ (S \circ e_B)^k \, .
\end{eqnarray*}
\end{itemize}

\subsection{Information Theory}

Recently a new application of domain theory has been introduced by Coecke and Martin \cite{CM02}. One of the main
results of their work is to show a domain formulation of the existing results from information theory. They have
shown that Shannon entropy and Von Neumann entropy can be captured as Scott continuous functions over the
corresponding domain. Here we briefly review their work in order to give a complete picture of quantum domain
theory. All the definitions and results in this subsection is taken from \cite{CM02}.

Coecke and Martin have constructed a domain structure over mixed states such that pure sates are the maximal
elements. They first order classical states recursively in terms of Bayesian order.

\begin{definition}
Let $n\geq 2$. The {\em classical states} are
\begin{eqnarray*}
\Delta^n = \left\{x\in[0,1]^n | \sum^{n}_{i=1}x_i =1 \right\}
\end{eqnarray*}
A classical state $x\in \Delta^n$ is {\em pure}, when $x_i=1$ for some $i\in{1,\cdots,n}$. The set of all pure
states is denoted by $\{e_i\,|\,i=1,\cdots,n\}$.
\end{definition}
A classical state in $x \in \Delta^n$ can be interpreted as the information that an observer has about the results
of an event in which $n$ different outcomes is possible, i.e. $x_i$ indicates the probability of obtaining the
outcome $i$. If we know $x$ and after measuring we determine that outcome $i$ is not possible, our knowledge
improves to
\begin{eqnarray*}
p_i(x)= \frac{1}{1-x_i}(x_1,\cdots,x_{i-1},x_{i+1},\cdots,x_n) \in \Delta^{n-1} \, ,
\end{eqnarray*}
where $p_i(x)$ is obtained first by removing $x_i$ from $x$ and then re-normalising. The partial functions $p_i$:
\begin{eqnarray*}
p_i:\Delta^{n}\rightharpoonup\Delta^{n-1} \, ,
\end{eqnarray*}
with $\dom(p_i)=\Delta^n\setminus{e_i}$, are called the {\em Bayesian projections}. The classical states are
partially ordered with the following recursive relation.

\begin{definition}
Assume that $x$ and $y$ are in $\Delta^n$ we write $x\po_B y$ iff:
\begin{eqnarray*}
\forall i : \;\; x,y \in \dom(p_i)\Rightarrow p_i(x) \po_B p_i(y) \, .
\end{eqnarray*}
For $x,y \in \Delta^2$ we have:
\begin{eqnarray*}
x\po_B y \;\; \mbox{iff}\;\; (y_1\leq x_1\leq 1/2) \; or \; (1/2\leq x_1 \leq y_1) \, .
\end{eqnarray*}
The above relation is called {\em Bayesian order}.
\end{definition}

The Bayesian order leads to a domain of classical states where the pure states are the maximal elements.

\begin{theorem} \cite{CM02} $(\Delta^n,\po_B)$ is a domain with the following set of maximal elements:
\begin{eqnarray*}
\{ e_i\,|\,1\leq i\leq n\} \, ,
\end{eqnarray*}
and least element $\bot=(1/n,\cdots,1/n)$.
\end{theorem}

Coecke and Martin have generalised the idea of Bayesian order to the quantum setting using the spectral order.
Informally speaking, to compare the amount of information of two given mixed states it is enough to consider an
observable and measure both mixed states. The result of measurements are two classical states and can be ordered
via Bayesian order. Following the notation of \cite{CM02}, we denote by $\Omega^n$ the set of all density matrices
on ${\cal H}^n$. For simplicity we also consider the following definition.

\begin{definition}
Assume that $O$ is a non-degenerate observable on ${\cal H}^n$ i.e. it has $n$ different eigenvalues with
orthogonal eigenvector spaces $\{P_i\}_{i=1}^{n}$. For a density matrix $\rho$ on ${\cal H}^n$ we define:
\begin{eqnarray*}
\spec(\rho|O)=(\tr(P_1 \cdot \rho),\cdots,\tr(P_n \cdot \rho)) \in \Delta^n \, .
\end{eqnarray*}
\end{definition}

\begin{definition}
Let $n\geq 2$, for quantum states $\rho,\sigma \in \Omega^n$, we have $\rho \po_S \sigma$ iff there exists a
non-degenerate observable $O:{\cal H}^n\rightarrow{\cal H}^n$ such that $[\rho,O]=[\sigma,O]=0$ and
\begin{eqnarray*}
\spec(\rho|O)\po_B \spec(\sigma|O) \, .
\end{eqnarray*}
This is called the {\em spectral order}.
\end{definition}

Finally the domain of the quantum states can be defined with:

\begin{theorem} \cite{CM02} $(\Omega^n,\po_S)$ is a domain with the following set of pure states as the maximal
elements and least element $\bot=I/n$, where $I$ is the identity matrix.
\end{theorem}

The final part of the Coecke and Martin's work that we review here is concerned with measuring of information
content \cite{CM02}.

\begin{definition}
A Scott continuous map $\mu$ from a given domain $D$ to the domain of \footnote{The set $[0,\infty)^*$ is the
domain of nonnegative real numbers in their opposite order.} $[0,\infty)^*$ {\em measure the content of} $x \in D$
if
\begin{eqnarray*}
x \in U \;\; \Rightarrow \;\; (\exists \epsilon > 0) \; x\in \mu_{\epsilon}(x)\subset U\, ,
\end{eqnarray*}
whenever $U$ is Scott open in $D$ and
\begin{eqnarray*}
\mu_{\epsilon}(x)= \{y\in D| y\po x \; \& \; |\mu(x) -\mu(y)| < \epsilon \}\, .
\end{eqnarray*}
The map $\mu$ {\em measures} $X$ if it measures the content of each $x \in X$.
\end{definition}

A map $\mu$ is a measure of content if it distinguishes the maximal (in content) elements.
\begin{definition}
Assume $D$ is a domain, a {\em measurement} is a Scott continuous map \\ $\mu : D \rightarrow [0,\infty)^*$ that
measures the set $\{ x \in D\,|\, \mu(x)=0\}$.
\end{definition}

The following results from \cite{CM02} present the domain picture of the well known functions, the Shannon entropy
and the von Neumann entropy.

\begin{theorem} \cite{CM02}
Shannon entropy
\begin{eqnarray*}
\mu (x) = - \sum_{i=1}^{n} x_i \log(x_i) \, ,
\end{eqnarray*}
is a measurement of type $\Delta^n \rightarrow [0,\infty)^*$.

von Neumann entropy
\begin{eqnarray*}
\sigma (\rho) = - \tr(\rho\lg(\rho) \, ,
\end{eqnarray*}
is a measurement of type $\Omega^n \rightarrow [0,\infty)^*$.

\end{theorem}

\section{Discussions}

In this paper, we have discussed a new framework for quantum computation via quantum domain theory. Using domain
theory a rigours framework for quantum computability has been also introduced. Although it is known that the class
of computable functions with a quantum model is equal to the class of classical computable functions
(Church-Turing Principle \cite{Deutsch85,Jozsa91,Ozawa98}), however we believe that by considering different
frameworks for quantum computability, interesting questions can be addressed. Quantum domain theory also provides
a topological structure for quantum computation that could be useful for the study in other fields of quantum
computation.

Furthermore we introduced a denotational semantics for quantum computation and we showed that quantum computation
over density matrices with completely positive maps, has a similar semantical structure as probabilistic
computation over random variables. This could be considered as a foundation for designing a functional programming
language for quantum computation.

Finally we briefly reviewed a domain structure for quantum information theory, introduced by Coecke and Martin
\cite{CM02}. The partial order introduced in their work has interesting connections with theory of entanglement as
they have discussed in their paper. Therefore, a domain theoretical approach to the theory of entanglement
manipulation may provide us with a uniform framework for measuring the entanglement in the same line as
\cite{VK02}.

\section*{Acknowledgement}

I would like to thank Angelo Carollo, Jens Eisert, Ivette Fuentes Guridi, Barry C. Sanders, Vlatko Vedral and
Herbert Wiklicky for useful comments and discussions on this topic.

\end{document}